%% file: spin-mu.tex
\documentclass[12pt,a4paper]{JHEP}
\usepackage{array,cite}
\usepackage{epsfig}
\input{defs}

\definmath\deltaetalep{\Delta\eta_{\ell^+\ell^-}}
\definmath\costhetall{\cos\theta^*_{ll}}
\definmath\ww{W^\pm W^\mp}
\definmath\wz{W^\pm Z^0}
\def\mess#1{{\typeout{AJB says: #1}}}
\title{
Measuring slepton spin at the LHC
}

\author{A.J. Barr\\
        Department of Physics \& Astronomy,
	University College London,
	Gower Street,
	LONDON,
	WC1E 6BT}

\abstract{
A new method is presented for measuring the spin of 
selectrons and smuons at the Large Hadron Collider (LHC),
using an angular variable which is sensitive to the 
polar angle in direct slepton pair production.
This variable, \costhetall, is invariant under boosts along the beam axis, 
so it can be used at the LHC despite the fact that
the longitudinal boost of the centre-of-mass frame cannot be determined.
Monte Carlo simulations demonstrate that, using this method,
the LHC can distinguish between the supersymmetric production 
angular distribution and phase space, 
or between supersymmetry and the production angular 
distribution of universal extra dimensions.
An integrated luminosity of about 100 to 300 \invfb\ provides sufficient 
statistics to measure the slepton spin for points 
which had left-handed slepton masses in the range 202 to~338 GeV, 
and right-handed sleptons in the range 143 to 252~GeV.
Good sensitivity was found in the `bulk' and `stau co-annihilation'
regions of the cMSSM favoured by cosmological relic density measurements.
Various systematic uncertainties are investigated, 
and some methods for reducing them are discussed.
}
\keywords{
Hadronic Collider, LHC, Supersymmetry, Spin, ATLAS, Universal Extra Dimensions
}
\preprint{
ATL-PHYS-PUB-2005-023
}

\begin{document}

\section{Introduction}

One of the main goals of the LHC is to search for new physics,
beyond the Standard Model (SM). In most cases, if new particles exist at 
the TeV-scale, then the LHC experiments will be able to discover their existence. 
But discovery is only the first step -  
to distinguish between models, the properties of the new particles 
also need to be experimentally determined.

Probably the leading contender for new physics at the TeV 
scale is supersymmetry (SUSY).
The minimal supersymmetric extension to the SM predicts that 
each SM particle should have a partner with
spin differing by \half. If R-parity is conserved, then the 
supersymmetric partners must be pair-produced and the lightest 
supersymmetric particle (LSP), typically the \ntlone, is stable. 
The typical LHC signature is therefore a pair of cascade decays 
which produce jets and leptons, as well as missing energy from 
the (assumed invisible) lightest supersymmetric particle.

There are, however many other competing models for new physics 
at the TeV scale besides supersymmetry.
Indeed it has recently been appreciated that  models with Universal Extra Dimensions 
(UED) and Kaluza-Klein (KK) parity\cite{Appelquist:2000nn}
could have very similar collider phenomenology to supersymmetric 
model\cite{Cheng:2002ab}.

The minimal version of UED predicts that for each SM particle there should be a 
tower of Kaluza-Klein (KK) excitations. 
KK parity means that particles with odd KK-number, 
such as the first excited state of any SM particle can only be produced in pairs.
It also ensures that the lightest KK particle must be stable, in the same way as R-parity
does for supersymmetry.
Distinguishing between SUSY and UED could therefore be a difficult problem, since
both models predict TeV-scale pair-produced particles 
which decay through cascades with Standard Model couplings, 
with the eventual production of a pair of invisible daughters (LSP or LKP).


While other measurements might be indicative\cite{Datta:2005zs,Datta:2005vx},
the property which will a give conclusive answer
as to whether we are observing SUSY or UED 
is the {\em spin} of the excited particles.

Recently some progress has been made towards spin-determination
of supersymmetric particles at the LHC.
The method, suggested in \cite{Barr:2004ze}
and investigated in \cite{Smillie:2005ar,Datta:2005zs,Goto:2004cp},
involved measurement of the lepton charge asymmetry
in $\ell q$ invariant mass distributions in the cascade decay,
\begin{equation}
\label{eq:partchain}
\sql \to \ntltwo\ q_L \to \sslrpm\ l^\mp\ q_L\ .
\end{equation}
That measurement was shown to have sensitivity to the spin of the \ntltwo.
While it was comforting to see that the LHC can have sensitivity to 
sparticle spins, the caveat is that 
in some parts of parameter space, the decay chain \eqref{eq:partchain} 
is kinematically forbidden or has a small branching ratio.
This makes it important to investigate {\em other} channels
and {\em other} particles for which the LHC experiments could measure spin.

In this paper we present a new method for measuring slepton spin
at the LHC.
The paper is organised as follows.
In \secref{sec:angular} we introduce an angular variable \costhetall,
and show that it is sensitive to the production polar angle in slepton 
pair production.
Our supersymmetric test points, Monte Carlo event generator and 
detector simulation are described in \secref{sec:mc}. 
In \secref{sec:cuts} we identify an event selection
and demonstrate that it can cleanly isolate the signal process. 
Results showing the experimentally-measurable angular distributions
and luminosity requirements are shown in \secref{sec:results}.
In \secref{sec:systematics} we discuss the main
systematic uncertainties and some methods for reducing them.
Our conclusions are presented in \secref{sec:conclusions}.

\section{Angular distributions, and \costhetall}
\label{sec:angular}
\EPSFIGURE{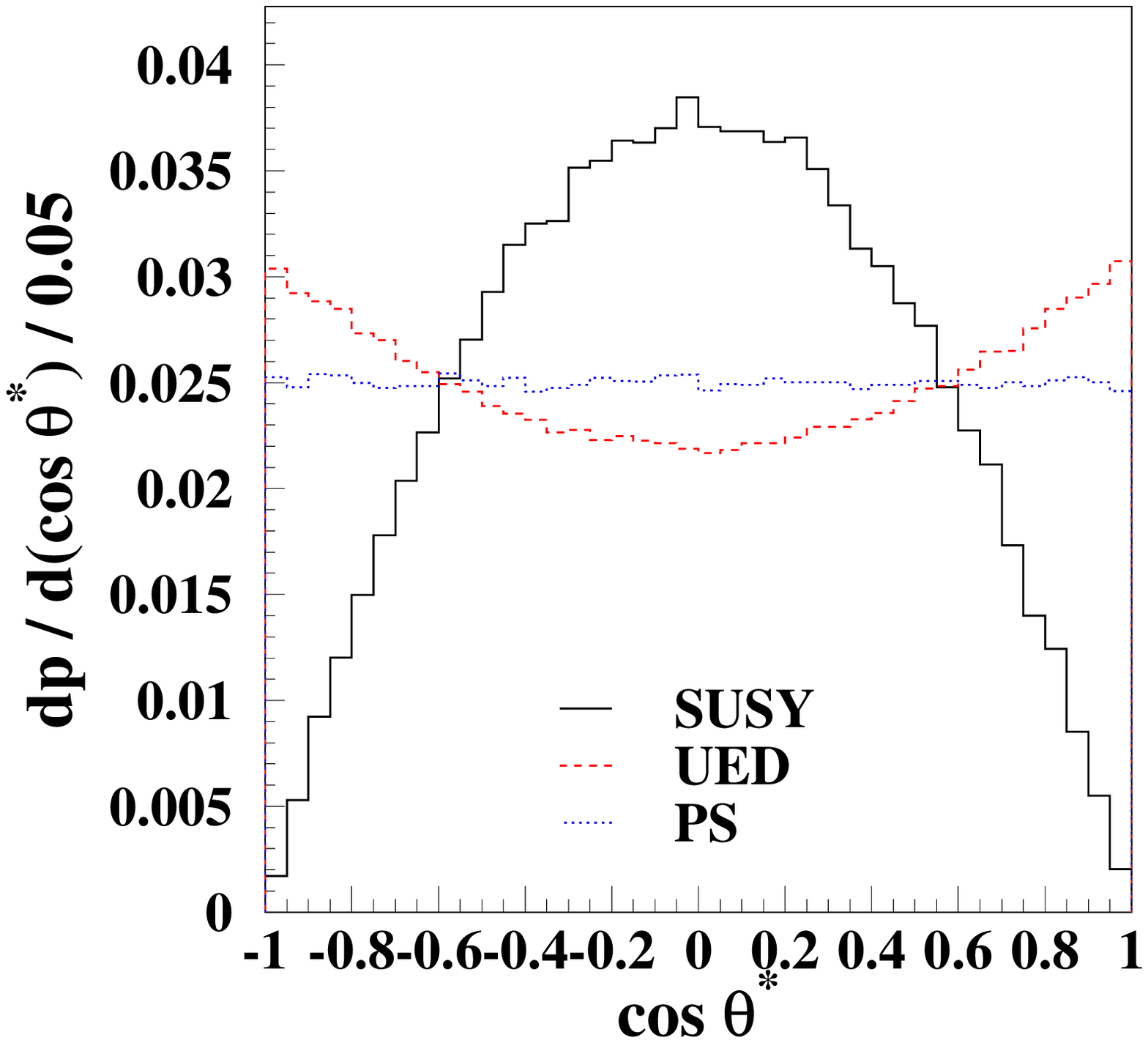, height=6cm, width=7cm}{
Production angular distributions, $\frac{dp}{d\cos\theta^*}$,
for scalar sleptons (SUSY), spin-\half KK leptons UED and pure phase space (PS).
The mass spectrum for the UED distribution is that of SUSY point S5 (see \secref{sec:mc}).
\label{fig:angles}
}

In this paper we investigate the supersymmetric process,
\begin{equation}
q\bar{q} \rightarrow Z^0/\gamma \rightarrow \tilde{\ell}^+\tilde{\ell}^- 
\rightarrow \ntlone \ell^+\ \ntlone \ell^- \ ,
\label{eq:slepton-decay}
\end{equation}
where throughout this paper $\ell$ is understood to mean electron or muon only.
Since sleptons are scalars, 
the angluar distribution for Drell-Yan slepton pair production is
\begin{equation}
\left(\frac{d\sigma}{d\cos \theta^*}\right)_\mathrm{SUSY} \propto 1-\cos^2\theta^*
\label{eq:slepton-dist}
\end{equation}
where $\theta^*$ is the angle between the incoming quark in one of the protons
and the produced slepton.
Slepton pair production via gauge boson fusion \cite{Choudhury:2003hq}
is not considered here, but it would become important for sleptons with masses
greater than about $300-400$~GeV.
For comparison we use a pure phase space distribution,
\begin{equation}
\left(\frac{d\sigma}{d\cos \theta^*}\right)_\mathrm{PS} \propto {\mathrm{constant}}\ .
\label{eq:ps-dist}
\end{equation}
The phase space distribution does not correspond to any physical model, but
does provide a convenient benchmark against which to compare the SUSY 
distribution.

We also compare against the UED equivalent of \eqref{eq:slepton-decay},
\begin{equation}
q\bar{q} \rightarrow Z^0/\gamma \rightarrow \ell_1^+\ell_1^- 
\rightarrow \gamma_1\ \ell^+ \gamma_1\ \ell^- \ .
\label{eq:kk-decay}
\end{equation}
which has the characteristic distribution for spin-\half\ KK leptons:
\begin{equation}
\left(\frac{d\sigma}{d\cos \theta^*}\right)_{\mathrm UED} 
\propto 1 + 
\left( \frac{E_{\ell_1}^2-M_{\ell_1}^2}{E_{\ell_1}^2+M_{\ell_1}^2} \right)\cos^2\theta^*\ ,
\label{eq:kklepton-dist}
\end{equation}
where $E_{\ell_1}$ and $M_{\ell_1}$ are the energy and mass respectively
of the KK leptons in the center-of-mass frame.
The three different production angular distributions are shown graphically
in \figref{fig:angles}.

\EPSFIGURE{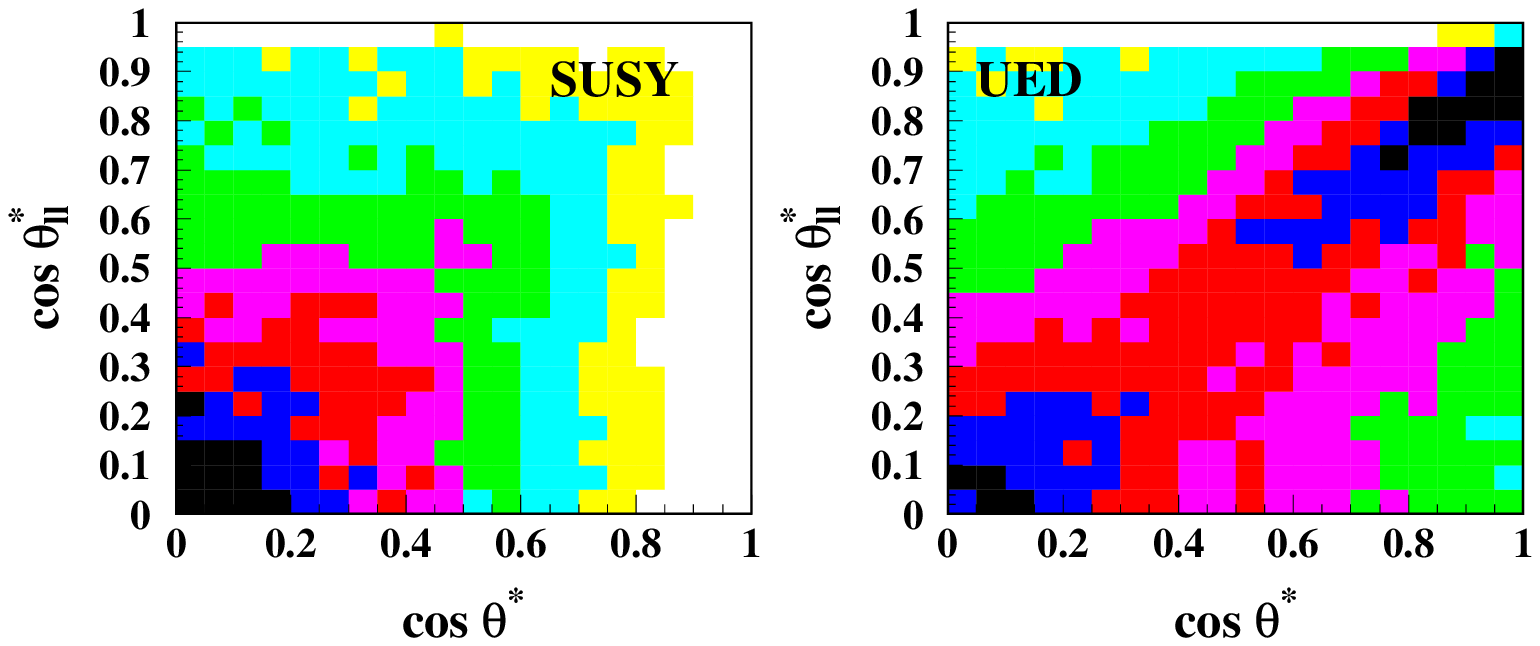, height=7cm}{
2-dimensional plots showing the correlation between our dilepton angular variable, 
\costhetall, (y-axes) and the cosine of the production angle of the 
parent sleptons  {\bf(a)} or KK-leptons {\bf(b)}
in the center of mass frame (x-axes).
Darker regions correspond to larger numbers of events,
the normalisation being arbitrary.
The mass spectrum is that of SUSY point {\tt S5}.
\label{fig:corr}
}

The different angular distributions provide a mechanism for determining
the heavy particle spin.
Excited leptons (selptons or KK-leptons) which are produced 
significantly above threshold will have decays which are boosted in the lab frame. 
This means that a pair of leptons from slepton decays 
(\eqref{eq:slepton-dist}) 
should be on average less widely separated in polar angle
than the pair from phase space (\eqref{eq:ps-dist}) or
KK-lepton pair production (\eqref{eq:kklepton-dist}).

It has already been suggested\cite{Battaglia:2005zf,Bhattacherjee:2005qe} 
that the final state lepton angular distributions 
could be used at a future high-energy $e^+ e^-$ linear collider
to distinguish between UED and SUSY models.
With a proton-proton collider such as the LHC, it is not possible to
measure the lepton angluar distributions in the 
parton-parton center-of-mass frame --
the initial $z$-momenta of the incoming partons are not known, and 
because invisible particles are produced, the center of mass frame of the 
parton-parton interaction cannot be recovered from the final state.

To make a spin measurement at a hadron collider, 
we propose a variable which is a function 
only of the pseudorapidity difference between the final state leptons, 
\deltaetalep. 
The advantage of differences in rapidity is that they are 
independant of the longitudinal boost.
The leptons are highly relativistic, so we can use their 
pseudorapidities as a very good approximation to their true rapidities.
By using a function only of \deltaetalep, we no longer have to face 
the problem of determining the center-of-mass frame along the beam direction.
The inter-lepton pseudorapdity difference, 
\deltaetalep, is also sensitive to the slepton
production angle. The reasons are the same reasons as for the 
lepton angular distributions -- 
the leptons `inherit' some knowledge of the rapidity
of their slepton or KK-lepton parents.
Lepton pairs from slepton pair decay will therefore 
be on average less separated in pseudorapidty
than those coming from particles produced according to 
the corresponding phase-space or Kaluza Klein production angular distributions.

To allow a more direct comparison with the production distributions, 
rather than using \deltaetalep\ directly, we propose the angular variable
\begin{equation}
\costhetall \equiv \cos{\left(2 \tan^{-1} \exp({\deltaetalep/2})\right)}\ = \tanh(\deltaetalep/2)\ .
\label{eq:newangledef}
\end{equation}
This variable, like \deltaetalep, has the benefit of being 
longitudinally boost invariant, but also has a simpler 
geometrical interpretation: \costhetall\ is the cosine of
the polar angle between each lepton
and the beam axis in the longitudinally boosted frame in which the 
(pseudo-)rapidities of the leptons are equal and opposite.

As can be seen from \figref{fig:corr}, 
\costhetall\ is indeed well correlated with the slepton or KK-lepton
production angle, $\cos\theta^*$.
The experimental observable \costhetall\ is on average smaller for 
SUSY (\figref{fig:corr}a) than for UED (\figref{fig:corr}b),
meaning that \costhetall\ can be usefully employed as a spin-sensitive
discriminant in slepton/KK-lepton pair production at hadron colliders.

\section{Monte Carlo simulation of test points}

\label{sec:mc}
\begin{table}[b]\begin{center}\begin{tabular}{|l|l|l|l|c|}
\hline
Point  & \ntlone & \sser  & \ssel & $\sigma(pp\to{\ell^+\ell^-})$\\
\hline
{\tt S5}     & 116   & 153    & 229 & 65~fb \\
\hline
{\tt SPS1a}  & 96    & 143    & 202 & 88~fb \\
\hline
{\tt SPS1b}  & 160   & 252    & 338 & 12~fb \\
\hline
{\tt SPS2}   & 80    & 1452   & 1456 & 0.003~fb \\
\hline
{\tt SPS3}   & 161   & 178    & 287 & 34~fb \\
\hline
{\tt SPS4}   & 119   & 417    & 448 & 2.5~fb \\
\hline
{\tt SPS5}   & 120   & 191    & 256 & 48~fb \\
\hline
{\tt SPS6}   & 189   & 237    & 265 & 22~fb \\
\hline
\end{tabular}\caption{
\label{tab:mass}
Masses of selected particles (in GeV) for the model points investigated,
and the sum of the leading order cross-sections for 
direct slepton $(e^+e^-~\mathrm{and}~\mu^+\mu^-)$ pair production.
}
\end{center}
\end{table}

We chose as a primary test point the LHC point 5\cite{phystdr}, 
({\tt S5}), with supersymmetric 
mass spectrum and branching ratios calculated with \isajetv{7.64}.
Further simulations were performed with the Snowmass points 
{\tt SPS1a}, {\tt SPS1b}, {\tt SPS2},  {\tt SPS3}, {\tt SPS4}, {\tt SPS5},  
and {\tt SPS6}\cite{Allanach:2002nj} with spectra and branching ratios calculated 
using \isajetv{7.58}, following the Snowmass standard.

Some important masses as well as the signal cross-section are given in \tabref{tab:mass}. 
It is immediately clear that {\tt SPS2}, which is in the `focus point'\footnote{
For a discussion of the various regions of cMSSM consistent 
with \ntlone\ dark matter see, for example, \cite{Ellis:2003cw}.
} 
region, and has very heavy sleptons, will not provide
a measurable signal at the LHC. Of the other points, we note that 
{\tt SPS4} has slepton masses near the edge of our expected sensitivity
for direct production\cite{Lytken:22}. The non-universal point {\tt SPS6} 
has a smaller-than-usual mass difference between the sleptons and the 
LSP, and so will produce rather soft leptons in its decays. 

The Monte Carlo event generator used was 
\herwigv{6.507}\cite{Corcella:2000bw,Moretti:2002eu,Richardson:2001df},
with the parton distributions of
{\tt MRST}\cite{Martin:1998sq} (average of central and higher gluon).
Since a complete UED Monte Carlo event generator programme was not available,
\herwig\ was modified to generate slepton pairs with 
the other production angular distributions, 
\eqref{eq:ps-dist} and \ref{eq:kklepton-dist}.
The slepton angular distribution (\eqref{eq:slepton-dist}) is of course 
already implemented in \herwig\cite{Moretti:2002eu}.
Angular distributions for UED processes other 
than for $\ell_1^+\ell_1^-$ production in \eqref{eq:kk-decay} 
were not modified. In general, this should not significantly 
affect the results presented, since cuts were later applied (\secref{sec:cuts}) which
reduced the residual supersymmetric background to a small fraction of the signal.

Unlike the case of slepton decay,
the extent to which the spin structure of the $\ell_1\to\gamma_1\ell$
vertex is important will depend on the details 
(such as the KK Weinberg angle  and left-right $\ell_1$ mixings) of the UED model.
We expect that others with an interest in particular 
UED models will investigate in more detail 
the performance of this method for those models.
For this paper, we merely note that the KK-leptons are only weakly polarised
(as can be seen from the similarity between the UED and PS distributions in 
\figref{fig:angles}), and so in our simple approximation to UED 
we ignore spin correlations in the $\ell_1$ decays.


Samples of inclusive supersymmetric particle production
were generated corresponding to integrated luminosities
from 25 to 500~\invfb\ for each of the test points.
Also, 500~\invfb\ samples were generated for the major backgrounds,
\ww, \zz, \wz, and \ttbar.
To understand the `theoretical' distributions, 
very large luminosity (10~\invab), samples were generated of the
signal process for each of the three production angular distributions
(\eqsref{eq:slepton-dist}, \ref{eq:ps-dist} and \ref{eq:kklepton-dist}),
and for each test point. In total about $8\times 10^8$ events were generated.

All of the events generated were passed through the ATLAS fast detector simulation
programme \atlfastv{2.50}\cite{Richter:1998at}, 
which contains a parameterisation of the 
ATLAS detector response for leptons, jets and missing energy.
Jets were defined by having 10~GeV of transverse energy in a cone of radius 0.4.
The pseudorapidity coverage for jets was $|\eta_j|<5$, while the 
lepton ($\ell\in e,\mu$) pseudorapidity coverage was $|\eta_\ell|<2.5$.
We assume a $b$-jet vertex tagging efficiency of 60\%, a $c$-jet 
vertex tagging rate of 10\% and light-quark jet mistagging rate of 1\%\cite{phystdr}.

\subsection{Event selection}
\label{sec:cuts}

The final state of interest consists of two opposite-sign, same family (OSSF) 
leptons, either electrons or muons, together with two invisible particles.
Thus the initial selection is a requirement that exactly two OSSF leptons
are detected, and that there must be missing transverse momentum, $\ptmiss > 100$~GeV.
The lepton with the larger transverse momentum must satisfy
$\pt(\ell^{(1)}) > 40$~GeV, 
while that with the smaller \pt\ must have $\pt(\ell^{(2)}) > 30$~GeV.
This lepton momentum selection also ensures that these events satisfy
the ATLAS trigger requrements.

The main Standard Model backgrounds are expected to come from
the final state $\ell^+\ell^- \nu\bar{\nu}$, from \ww\ or \zz\
production. Since \wz\ and \ttbar\ might also be significant, they 
were also generated.

The \zz\ backgrounds is largely removed 
by excluding events with dilepton invariant mass $m_{\ell\ell}<150$~GeV.

The \ww\ pair background can be substantially reduced by making a cut 
on a different kinematic variable, known as $\mttwo$.
This variable, described in \cite{Lester:1999tx,Barr:2003rg},
is similar to the transverse mass, $m_T$,
but is useful in events in which a {\em pair} of same-mass particles 
decays semi-invisibly.
It is defined by:
\begin{eqnarray}
{\mttwo}^2 (m_\nu) 
&\equiv& 
{ \min_{\slashchar{{\bf q}}_T^{(1)} +
\slashchar{{\bf q}}_T^{(2)} = {\bf \pmiss}_T }} {\Bigl[ \max{ \Bigl\{
m_T^2({\bf p}_T^{\ell^{(1)}}, \slashchar{{\bf q}}_T^{(1)}; m_\nu) ,\
m_T^2({\bf p}_T^{\ell^{(2)}} , \slashchar{{\bf q}}_T^{(2)}; m_\nu) \Bigr\} }
\Bigr]}.\label{eq:mt2}
\end{eqnarray}
where,
\begin{equation}
m_T^2 ( {\bf p}^{\ell}_T, {\bf p}_T^\nu; m_\nu) \equiv { m_{\ell^+}^
2 + m_\nu^2 +
2 ( E_T^{\ell} E_T^\nu - {\bf p}_T^{\ell}\cdot{\bf p}_T^\nu ) }\ ,
\label{MT2:MTDEF}\end{equation}
\begin{equation}E_T^{\ell} = 
{ \sqrt { ({\bf p}_T^{\ell}) {^2} + m_\ell^2 }}
\qquad \hbox{ and } \qquad {E_T^\nu} = 
{ \sqrt{ ({\bf p}_T^\nu){^2} + m_\nu^2 } }\ .
\label{MT2:ETDEF}
\end{equation}
and ${\bf p}_T^\ell$ and ${\bf p}_T^\nu$, indicate the 
lepton and neutrino transverse momenta respectively\footnote{
The neutrino mass is not important in this analysis, 
but is included in \eqsref{eq:mt2}-\ref{MT2:ETDEF} for completeness.}.
\mttwo\ has the property by construction that for events
in which the lepton pair originates from on-mass-shell \ww\ pairs,
\begin{equation}
\mttwo({\bf p}_T^{\ell^+},{\bf p}_T^{\ell^-}; m_\nu) < m_W\ .
\label{eq:mt2limit}
\end{equation}
We therefore exclude from our analysis any event for which 
$\mttwo({\bf p}_T^{\ell^+},{\bf p}_T^{\ell^-};0)<100$~GeV,
removing most of the \ww\ background.


To reduce the background from \ttbar\  production and from 
cascade decays of heavier KK or SUSY particles, 
the following additional selection was applied:
\begin{itemize}
\item{no jet with $\Pt>100$~GeV}
\item{no jet with a vertex tag (no `$b$-jets')}
\item{transverse recoil, $|\Ptmiss + \Pt(\ell^+) + \Pt(\ell^-)| < 100$}~GeV,
\end{itemize}
where the jet and vertex tagging algorithm parameters are as described previously.

Another potential background comes from single $W^\pm$ production 
in association with a jet or photon which fakes a lepton in the detector. 
The evaluation of such backgrounds would require a complete detector description, 
beyond the scope of this paper, and so they have not been included in this analyis. 
In fact we expect that although they have a much higher cross-section than
gauge boson pair production, they should not pose a serious threat.
The expected rate of jets faking electrons is rather small -- 
of the order $10^{-4}-10^{-5}$\cite{phystdr}. What is more,
these types of event should also be efficiently removed by the cuts.
Fake events where the $W^\pm$ is close to its mass shell and 
in addition either:
\begin{itemize} 
\item{there is no significant missing transverse momentum (\Ptmiss) except that 
caused by the neutrino from the $W^\pm$ decay {\em or}}
\item{any additional contribution to \Ptmiss\ is collinear with the `fake' lepton, 
as could be expected from a poor energy measurement of that fake,}
\end{itemize}
will fail the \mttwo\ cut.
The reason can be understood when one considers the quantity being minimized in \eqref{eq:mt2}
with a particular pair of value for the 2-vectors $\slashchar{{\bf q}}_T^{(1)}$ and $\slashchar{{\bf q}}_T^{(2)}$
which represent a possible transverse missing energy spitting. 
The interesting configuration is one in which $\slashchar{{\bf q}}_T^{(1)}$ is equal to the true
neutrino transvers momentum, and $\slashchar{{\bf q}}_T^{(2)}$ is parallel to the 
fake lepton. Under the conditions above, this configuration will satisfy the constraint 
${{\bf q}}_T^{(1)}+ {{\bf q}}_T^{(2)} = \Ptmiss$. 
The first transeverse mass, $m_T^{(1)}$, containing the neutrino and 
lepton from the $W^\pm$ decay, must be less than $m_W$, while
the other, $m_T^{(2)}$, containing the fake lepton, will be close to zero.
Since such a configuration exists, it places an upper limit on $\mttwo$ of  $m_W$.

\EPSFIGURE{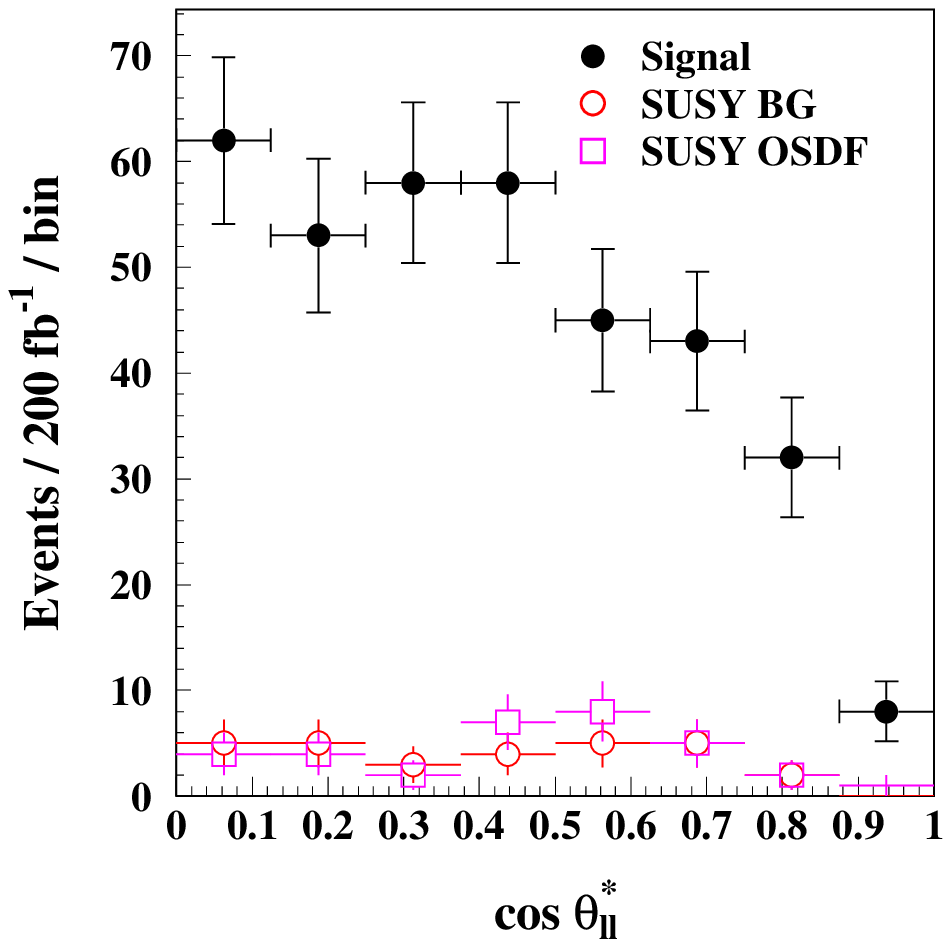, height=9cm}{
Distribution of \costhetall\ for the {\tt S5}
SUSY sample after event selection.
The dark cicles represents the SUSY slepton pair signal (\eqref{eq:slepton-decay}).
The open cicles indicate the SUSY background.
The open squares show the different flavour (OSDF) estimate of the SUSY 
background, which is explained in \secref{sec:sys:susy}.
In this and in all subsequent plots, the electron-pair and
muon-pair distributions have been combined.
\label{fig:s5plot}
}

\EPSFIGURE{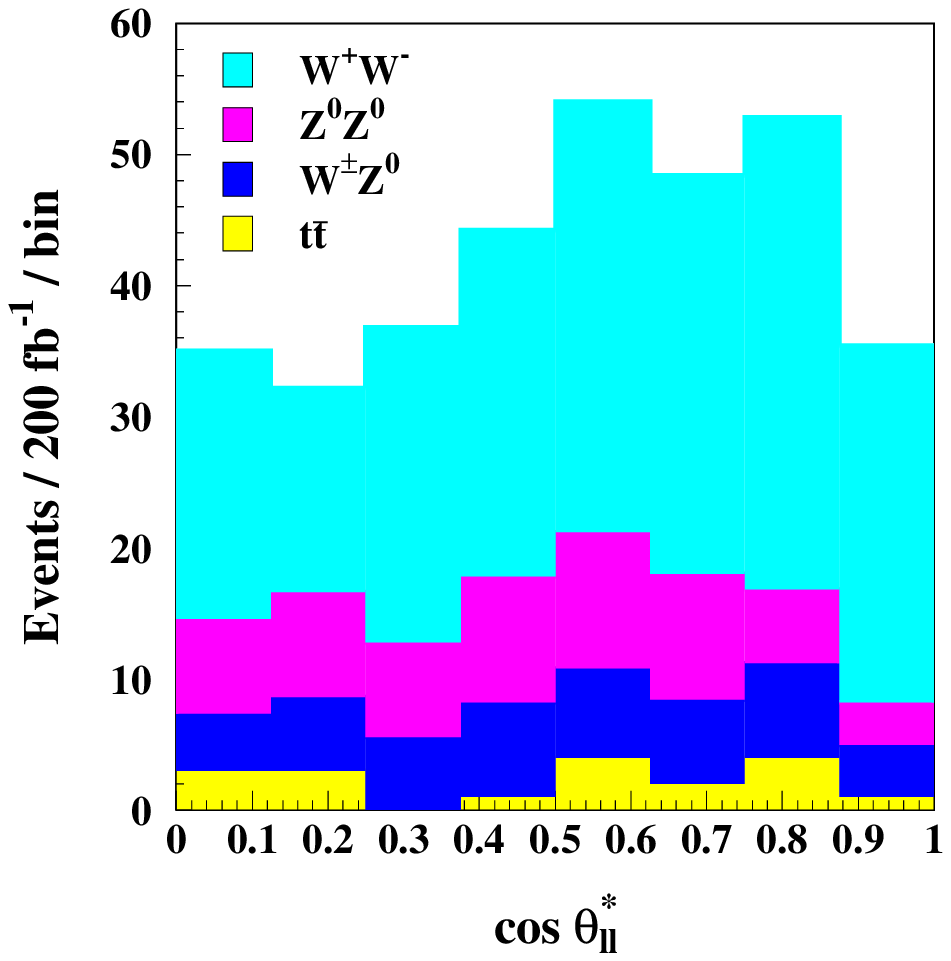, height=10cm}{
The Standard Model backgrounds after the cuts have been applied,
as a function of \costhetall.
\label{fig:sm-bg}
}

The cuts are somewhat similar to those used for studies of
slepton discovery potential, or mass measurements 
from Drell-Yan pair production at the 
LHC\cite{delAguila:1990yw,Baer:1993ew,Baer:1995va,
Lester:2001zx,Lytken:22,Weiglein:2004hn}.
One notable difference is that, unlike most previous studies,
we make no azimuthal angle ($\phi$) requirements on the leptons or \Ptmiss\ vector, 
since this seem to be unnecessary when an \mttwo\ cut is used to 
reduce the \ww\ background.


\Figref{fig:s5plot} shows that after the cuts have been applied, the
SUSY background is very much smaller than the signal. Similarly 
\figref{fig:sm-bg} demonstrates that the Standard Model
backgrounds do not overwhelm the SUSY signal. However the SM 
background is not insignificant, and so it will have to be
accurately determined if a slepton spin measurement 
based on \costhetall\ is to be achieved. 
The degree to which these backgrounds can be understood is investigated further in 
 \secsref{sec:sys:smbg} and \ref{sec:sys:susybg}.

\subsection{Results}
\label{sec:results}

In this section we disuss the \costhetall\ distributions which would 
be obtained assuming that the supersymmetric and 
Standard Model backgrounds can be accurately determined.
We then go on to examine the extent to which these assumptions 
can be justified in \secref{sec:systematics}.

\mess{About to do the S5 plot}

\EPSFIGURE{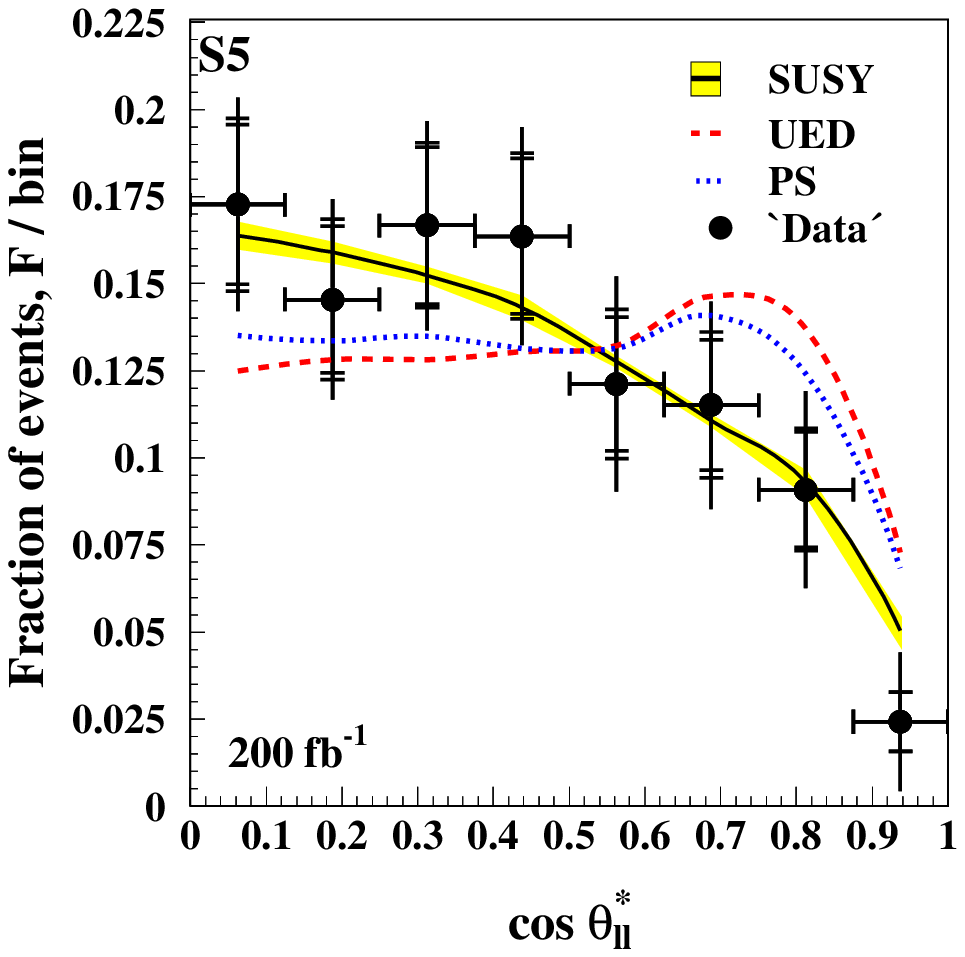, height=10cm}{
The points show the \costhetall\ distribution for 
the {\tt S5} signal sample ($ \tilde{\ell}^+\tilde{\ell}^- 
\rightarrow \ntlone \ell^+\ \ntlone \ell^- $)
after an integrated luminosity of $200~\invfb$.
The lines show the predictions for angular distributions 
according to supersymmetry (solid black line, \eqref{eq:slepton-dist}), 
phase space (dotted blue line, \eqref{eq:ps-dist}), 
and universal extra dimensions (dashed red line, \eqref{eq:kklepton-dist}). 
The error bars on the data show the statistical uncerainty on:
{\em inner error bar:} SUSY signal only;
{\em intermediate error bar:} inclusive SUSY with the
SUSY background subtracted;
{\em outer error:} inclusive SUSY with
both the SUSY and the SM backgrounds subtracted.
The narrow shaded band around the SUSY expectation 
shows how it is modified when the sparticle masses are 
simultaneously changed for all
sparticles by $\pm20$~GeV, as described in \secref{sec:sys:susy}.
Systematic uncertainties in the SUSY and SM background subtraction
are not included here, but are discussed in 
\secsref{sec:sys:smbg} and \ref{sec:sys:susybg}.
\label{fig:s5comparison}
}

\mess{Done the S5 plot}

In \figref{fig:s5comparison} we plot the fraction of events
as a function of \costhetall\ for $200~\invfb$ of integrated luminosity.
As has already been shown in \figsref{fig:sm-bg} and \ref{fig:s5comparison},
for point {\tt S5} the statistical uncertainty on the SUSY signal 
and the SM background are comparable, while the SUSY background is small.
It is clear that the (SUSY) ``data'' sample is much better
matched to the slepton angular distribution than to either the phase-space one
or the UED-like one. This means \costhetall\ does indeed 
measure the spin of the sleptons for this point.

\mess{about to do signif}

\EPSFIGURE{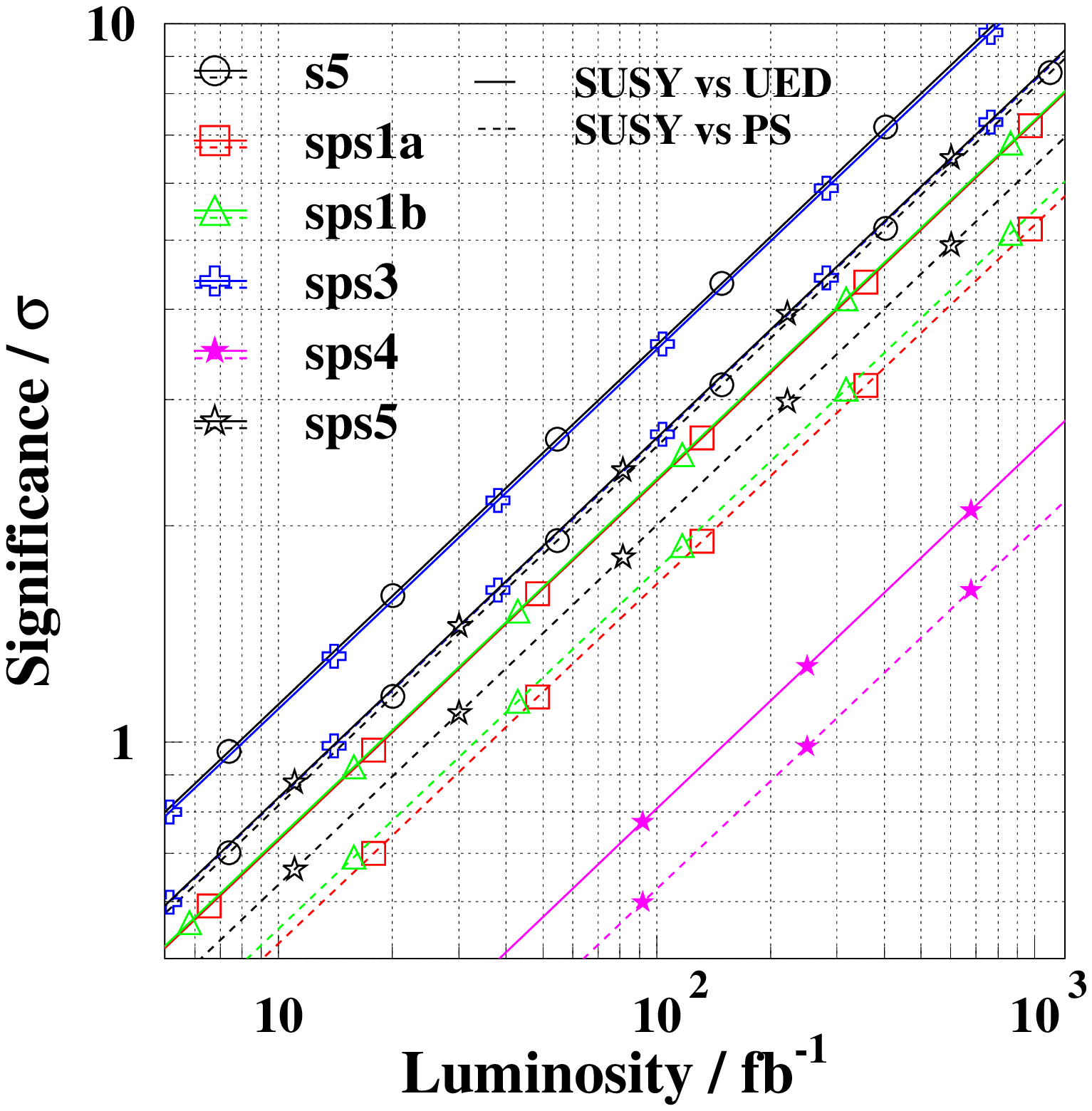, height=10cm}{
The expected statistical significance for discriminating 
SUSY from UED (solid lines), or SUSY from phase space (dashed lines) for
our SUSY test points, as a function of integrated luminsity.
Statistical uncertainties are included for the signal,
the SUSY background from slepton-pair to gaugino (other than \ntlone)
and the Standard Model backgrounds in \figref{fig:sm-bg}.
For each test point the the solid line (SUSY vs UED) always
requires lower luminosity.
\label{fig:signif}
}

\mess{Done the signif plot}

\begin{figure}
\begin{center}
  \begin{minipage}[b]{.49\linewidth}
   \begin{center}
    \epsfig{file=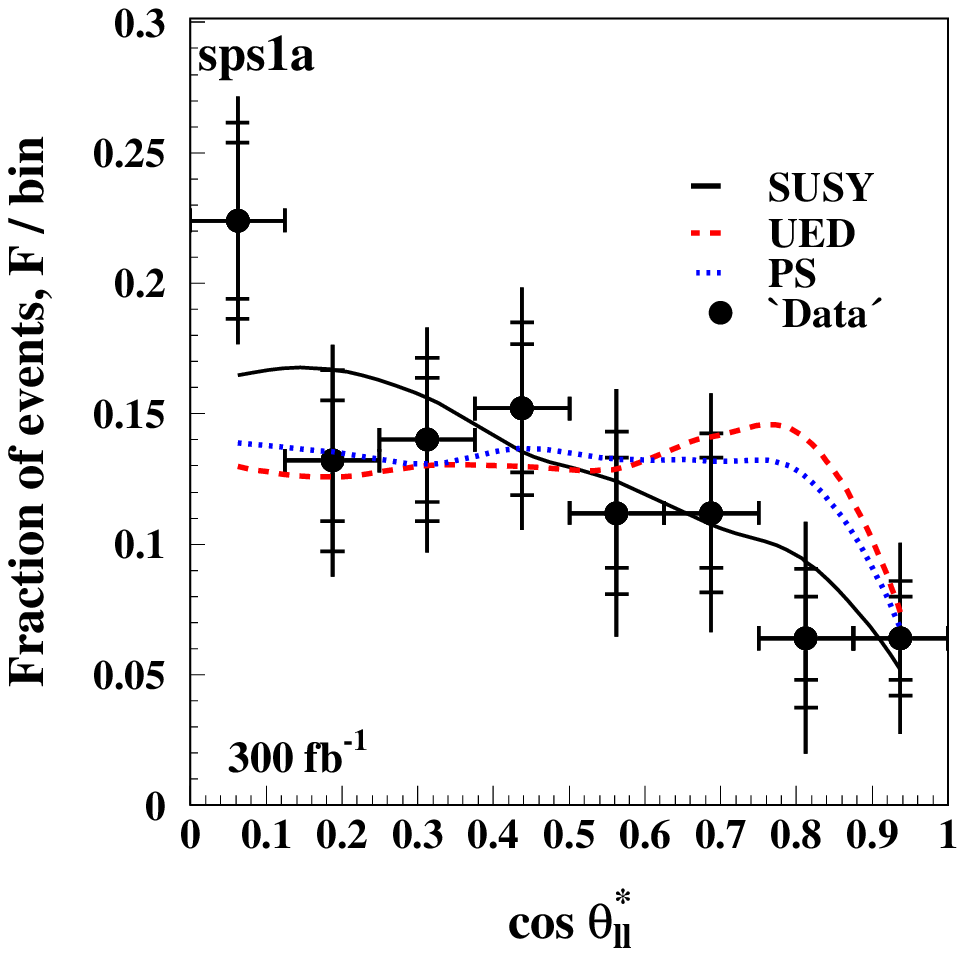, height=8.5cm}
     \\ \ \hspace{1cm} {\bf(a)}
    \epsfig{file=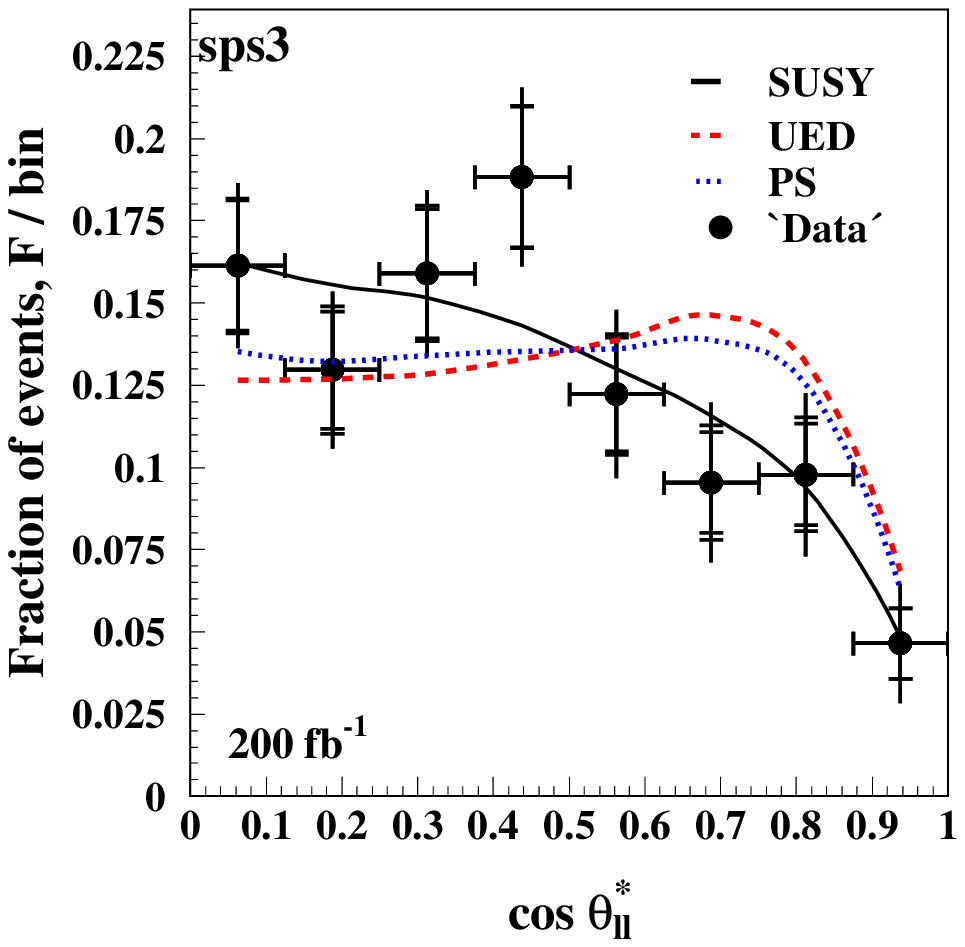, height=8.5cm}
     \\ \ \hspace{1cm} {\bf(c)}
    \end{center}
  \end{minipage}\hfill
  \begin{minipage}[b]{.49\linewidth}
    \begin{center}
    \epsfig{file=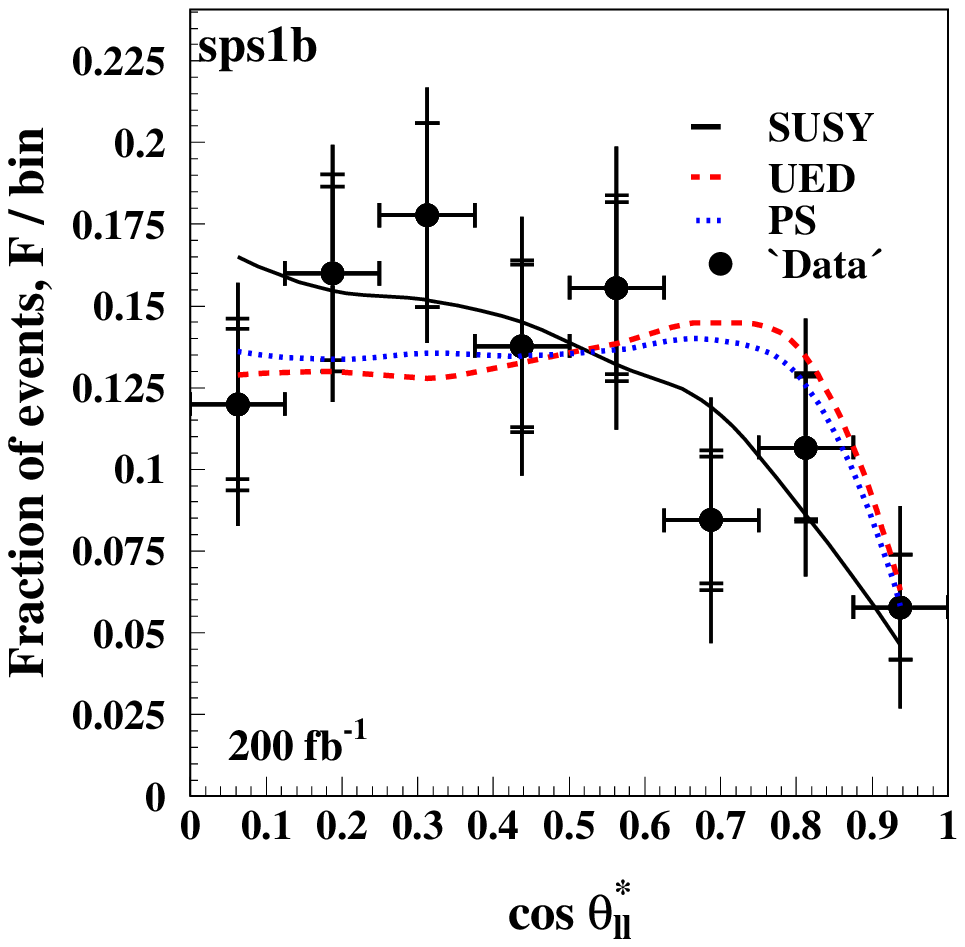, height=8.5cm}
     \\ \ \hspace{1cm} {\bf(b)}
    \epsfig{file=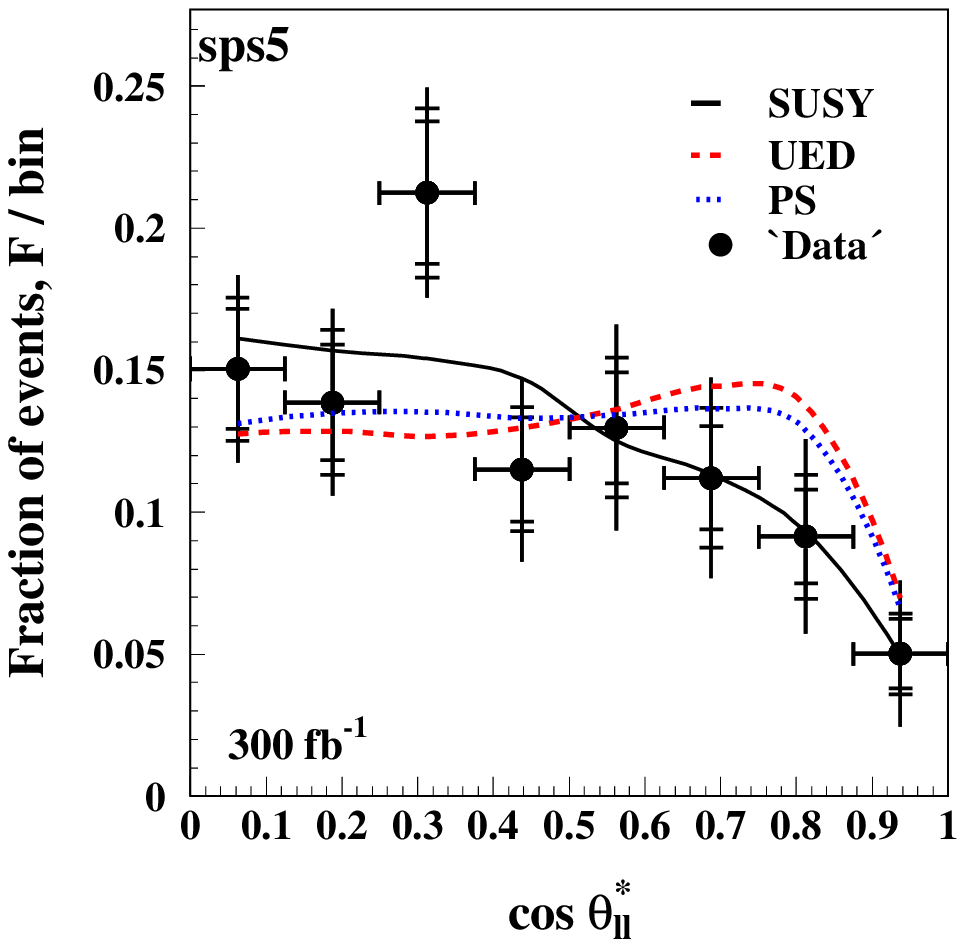, height=8.5cm}
     \\ \ \hspace{1cm} {\bf(d)}
    \end{center}
  \end{minipage}\hfill
\end{center}
\caption{
As for \figref{fig:s5comparison} but for Snowmass points
{\bf (a)} {\tt SPS1a}, {\bf(b)} {\tt SPS1b} 
{\bf (c)} {\tt SPS3} and {\bf(d)} {\tt SPS5}.
The integrated luminosity simulated in these plots is
200~\invfb\ for {\tt SPS1b} and {\tt SPS 3},
and 300~\invfb\ for {\tt SPS 1a} and {\tt SPS 5}.
\label{fig:many-plots}
}
\end{figure}

\mess{Done the many plots}

In \figref{fig:signif} we present the statistical separation expected 
for our test points  ({\tt S5} and the  
Snowmass points) as a function of integrated luminosity. 
The significance indicated is shows the gaussian-equivalent significance of
each of two tests:
\begin{enumerate}
\item{A test comparing the SUSY angular distribution (\eqref{eq:slepton-dist})
to the phase space one (\eqref{eq:ps-dist}) -- demonstrating that 
there is sensitivity to spin in the dynamics;}

and separately,
\item{A test comparing the SUSY angular distribution to the UED-like one 
(\eqref{eq:kklepton-dist})
-- showing discrimination between two physically-motivated models.
}
\end{enumerate}
In both cases we have used a discriminant which is
symmetrical in the hypotheses under test. The discriminant
accounts for the fact that, because the normalisation of the distribution 
is fixed, the data in the bins are correlated.
The events were counted in large statistics samples
after the cuts had been applied
and include the statistical uncertainty from the SUSY 
and SM background and the slepton to gaugino (other than \ntlone)
background subtraction, but none of the other systematic uncertainties 
discussed in \secref{sec:systematics}. 

Starting with the worst case, 
{\tt SPS2} (from the cosmological `focus point' region) 
has $>~1$~TeV mass sleptons, and such a small cross-section
for direct slepton production that no spin measurement 
(or indeed any other direct slepton production measurement) 
is possible for this point.
It is not included in \figref{fig:signif}.

{\tt SPS6} (also not on \figref{fig:signif}) also presents a difficult experimental case,
but for a quite different reason.
This point has non-universal gaugino masses, with the \ntlone\ 
more massive than would be the case with universal gauginos. 
Despite its large cross-section, a spin determination is very difficult
because the sleptons are only about 50, and 75~GeV heavier than the LSP, 
and the low-\pt\ leptons they produce end up buried beneath the $\ww$ background.

{\tt SPS4} has $\tan\beta=50$, 
rather heavy $\tilde{e}$ and $\tilde{\mu}$ sleptons (\tabref{tab:mass})
but relatively light \chgone\ and \ntltwo\ at 218~GeV.
The effect of this is that the signal process \eqref{eq:slepton-decay}
has to compete with cascade decays through the \chgone\ or \ntltwo\ 
and so the {\em supersymmetric} background is rather large at this point.
For this point, either very high statistics (such as could be achieved 
with an LHC luminosity upgrade\cite{Gianotti:2002xx}) or perhaps a very careful 
background reduction analysis, might allow the spin determination 
to be made in this channel.

The remaining five points, 
{\tt S5}, {\tt SPS1a},  {\tt SPS1b},  {\tt SPS3}, and {\tt SPS5}
all allow spin determination with $100-300~\invfb$ 
of integrated luminosity.
Reference distributions, similar to 
\figref{fig:s5comparison}, but for {\tt SPS} points,
are shown in \figref{fig:many-plots}.

{\tt S5} and {\tt SPS1b} which have masses `typical' of the SUSY 
`bulk' region require integrated luminosity of about 
100 and 300~\invfb\ respectively.
The point {\tt SPS5}, which is distinguished by a very light stop 
squark, 220 GeV,
behaves in this analysis like any other `typical' spectrum.
{\tt SPS1a} requires a slightly higher than expected integrated luminosity
(about 300~\invfb) despite it having the largest production cross-section 
(\tabref{tab:mass}). 
This is because its lighter sleptons produce relatively soft leptons which  
more frequently fail the cuts, and so the cross-section {\em after} selection
is actually smaller than for many of the other points.

Point {\tt SPS3} is in the cosmological `stau co-annihilation' 
region, in which the lighter stau is nearly degenerate with the
\ntlone. One might worry that the leptons it would produce might be too 
soft for this analysis. While this is true for the right-handed sleptons,
which are only 17 GeV heavier than the LSP,
a good signal is found from the heavier left-handed sleptons, 
and it turns out that 
this point is in fact one of the easier ones with which 
to make this spin measurement.

In one sense \figref{fig:signif} is optimistic, in that it
does not correctly take account of all the experimental systematic
uncertainties explored in \secref{sec:systematics} -- only the
statistical effect of the SUSY signal, and the 
SUSY di-slepton to gaugino and Standard Model background subtractions
are included.
In another sense this plot is conservative, as 
{\em no tuning} of the cuts has been done between the different points.
With the real data, the cuts could be tuned according to the masses 
of the particles already measured.
Such tunings have not been attempted here but have previously been shown to provide 
somewhat improved results for making mass measurements
at different slepton masses\cite{Baer:1995va,Lytken:22}.

Overall, we can see that, unlike the previous 
lepton charge asymmetry measurement\cite{Barr:2004ze},
our new method appears to be rather general.
There are some difficult cases, when the sleptons are heavy
or the slepton--LSP mass difference is small. 
However spin determination is reachable with statistics
corresponding to a few years of LHC design luminosity for 
a variety of points coming from both the `bulk' and the 
`stau co-annihilation' regions favoured by cosmology.

\section{Systematic uncertainties}
\label{sec:systematics}

It is important that neither the experimental data
points nor the comparison distributions are biased in a way which
cannot be either measured or calculated. 
Having shown in \secref{sec:mc} that the 
spin measurement is within statistical reach,
we now address the question of how the various 
systematic uncertainties can be controlled.

\subsection{Lepton efficiency, acceptance and charge identification}
\label{sec:sys:effacc}

\EPSFIGURE{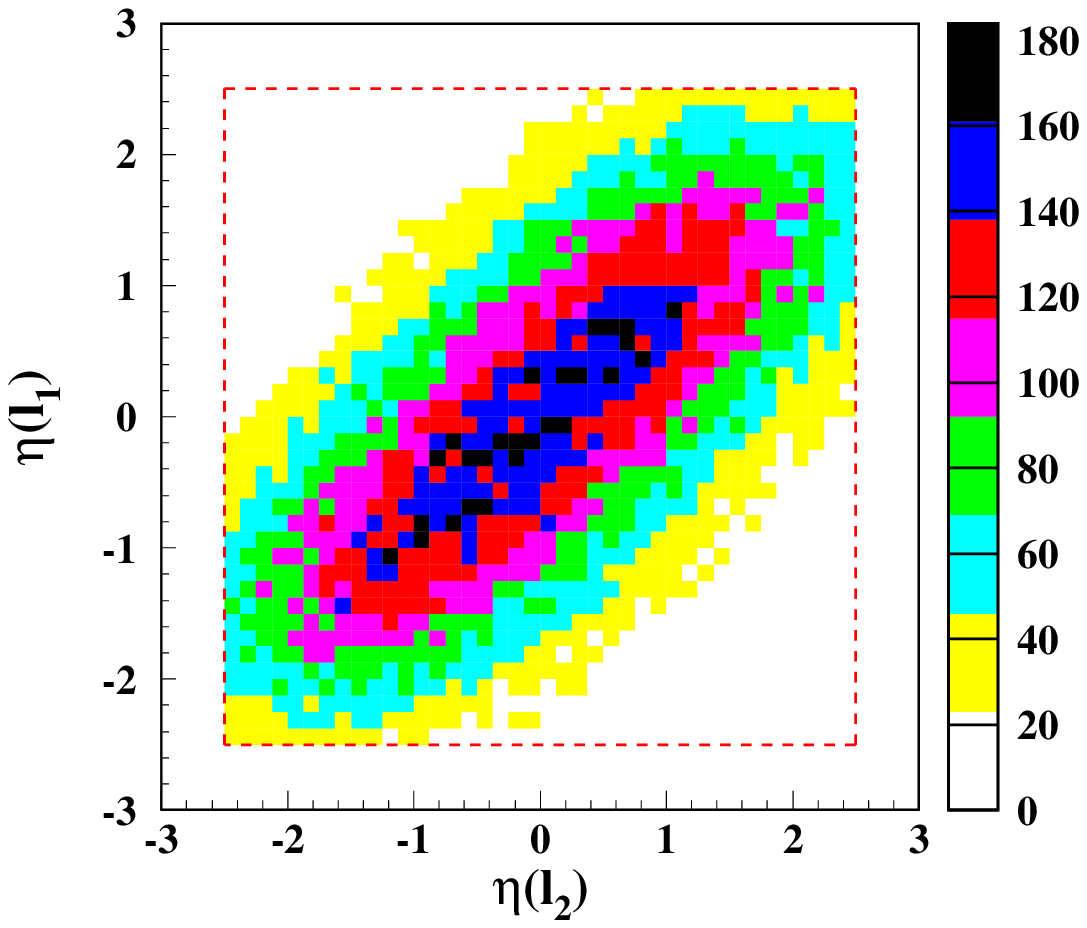, height=9cm}{
Reconstructed pseudorapidity ($\eta_1,\eta_2$) 
of the each of the two leptons in a large sample of signal events at SUSY point {\tt S5}. 
The \atlfast\ lepton acceptance limits at $|\eta|=2.5$ are indicated by the 
dashed lines.
\label{fig:acceptance}
}
\EPSFIGURE{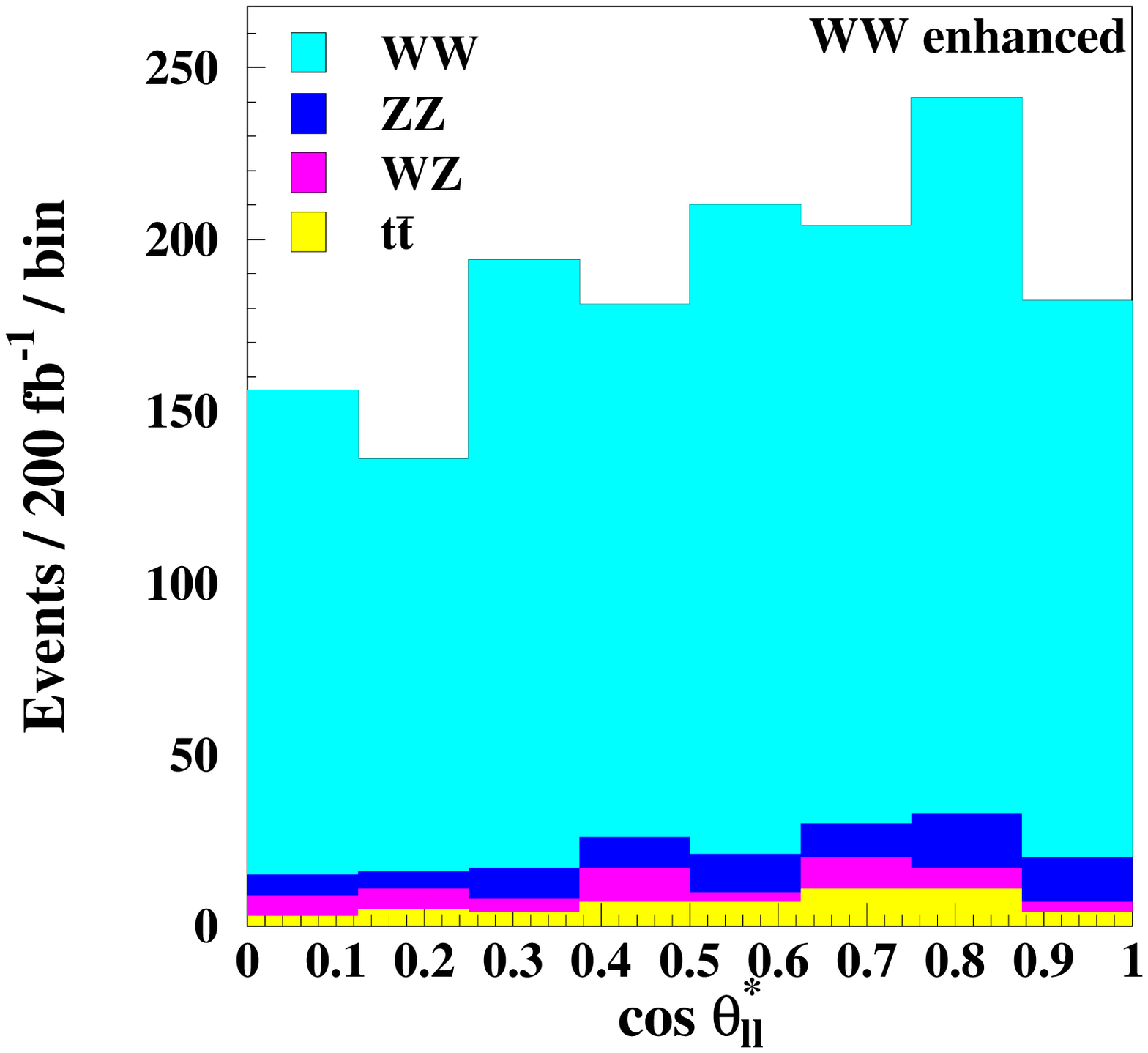, height=9cm}{
$\costhetall$ distribution for a control sample in which the 
$W^\pm W^\mp$ pair contribution has been enhanced by reversing the cut on \mttwo.
\label{fig:control}
}

Since the sensitive variable is based on \deltaetalep, 
it is clear that the electron and muon reconstruction efficiency 
will have to be well known across the range of pseudorapidity.
In ATLAS, the reconstruction efficiency is expected to be accurately determined 
from the experimental data, using the high statistics sample of 
$Z^0 \rightarrow \ell^+\ell^-$ available at the LHC. 
One of the leptons is used as a tag 
and the efficiency is determined from the fraction of events in which 
the second lepton is identified.
The same sample, and the same methods can be used for determining
the charge mis-identification rate. This can be done for different
values of $\eta$, and so the reconstruction efficiency should be
well under control for all values of $\eta$, including around 
the barrel-endcap transition, and near the limits of the detector acceptance.

Focusing our attention more closely on the edge of the
detector acceptance, we note that 
it is partly due loss of events beyond acceptance
that there is a down-turn in the \costhetall\ distributions 
(\figsref{fig:s5comparison} and \ref{fig:many-plots}) 
near unity. To investigate the sensitivity to the 
experimentally-forced rapidity selection we plot in
\figref{fig:acceptance}, the pseudorapidity distributions of the 
selected leptons.
Only a small proportion of entries occur near around ($|\eta|=2.5$),
suggesting that uncertainties in the acceptance should not cause a 
significant bias in the results.

\subsection{Standard Model background determination}
\label{sec:sys:smbg}

The normalisation for each Standard Model background can be 
determined independantly from the data,
by measuring the cross-section in regions of phase space in which 
each particular background dominates. 
For example, if one reverses the \mttwo\ cut, so that {\em only} events
with $\mttwo<100$~GeV are {\em accepted}, one obtains a sample of
events rich in \ww\ boson pairs (\figref{fig:control}).
In the same spirit, 
it is not difficult to define cuts which preferentially select
$Z^0$ boson pairs, \wz\ events or \ttbar\ pairs.
Some care obviously needs to be taken to ensure that a bias is not introduced
when extrapolating from one kinematical region to another. 
The experimental techniques required for such an extrapolation are 
not investigated here, but will be the subject of a future study\cite{tomalan}.
Previous work on SM background subtraction
using the flavour subtraction methods discussed in 
\secref{sec:sys:susybg} has already shown good promise\cite{Lytken:22}.

\subsection{SUSY backgrounds}
\label{sec:sys:susybg}

Although the SUSY backgrounds are generally small 
compared to the signal (\figref{fig:s5plot}), they do need to be understood
so that they can either be confirmed to be
too small to worry about, or, more realistically, so
that they can be accurately determined and subtracted.

Frequently-used methods\cite{Baer:1995va,Goto:2004cp}
for measuring the background rate from the data 
are based on constructing the equivalent distribution as for the signal, 
but rather than requiring an opposite sign, same family (OSSF) dilepton pair,
asking instead for a same sign, same family (SSSF e.g. $=e^+,e^+$) pair; 
or an opposite sign, different family (OSDF, e.g. $\mu^+e^-$) pair; 
or a same sign, different family (SSDF e.g. $e^+\mu^+$) pair. 
Combinations of SSSF, OSDF, and SSDF selections can be employed
along with particular background-enhanced samples to estimate the various backgrounds. 

For high-scale processes at the LHC a flavour subtraction rather 
than a charge subtraction method is generally most powerful.
Lepton universality is only broken by the difference
between the $\mu$ and the $e$ Yukawa couplings, so distributions involving
hard, isolated leptons (as opposed to those coming from near-threshold or Dalitz decays) 
can be generally expected to be rather universal between generations.
By contrast at high scales we are sampling the proton's parton distribution
functions in the valence region, so the initial state is not charge symmetric, 
and so neither are the final states, 
reducing the effectiveness of a charge-subtraction method.

For slepton pair production most of the SUSY backgrounds can be expected to 
have similar cross-sections in the OSSF (signal)
and the OSDF (control). 
The good similarity between the SUSY background and an OSDF selection
for our {\tt S5} point can be seen in \figref{fig:s5plot}.
It has been shown \cite{Lytken:22} that a simple 
OSSF-OSDF subtraction (without using control samples) 
can provide a rather clean background subtraction
for both the SUSY background and most of the Standard Model backgrounds.

Returning for a moment to the SM backgrounds discussed in \secref{sec:sys:smbg},
the size of the \ww, \wz, \ttbar, and $W^\pm\ +$~`fake lepton'
can also be estimated with a flavour subtraction.
It is likely that the best results for both the SUSY and the SM background
determinations will come from a mixture of techniques, both 
of OSDF subtraction and control samples.

\subsection{SUSY spectrum}
\label{sec:sys:susy}

Another important issue to address is to what extent
our lack of knowledge of the {\em masses} of the SUSY particles
might lead to uncertainties in the comparison 
(`theoretical') distributions.
It is safe to assume that because of the higher integrated luminosity required, 
any spin measurement will be made {\em after} some knowledge of the 
sparticles masses has been obtained.
It is therefore sufficient to investigate
the effect on the comparison distributions caused by varying 
the sparticle masses within the limits which 
are expected after 100~\invfb\ or more has been collected.

Studies\footnote{Examples include \cite{phystdr,Tovey:2000wk,Lester:2001zx,Kawagoe:2004rz,Lester:2005je}.} suggest that 
with around 100~\invfb\ of integrated luminosity the 
LHC experiments can strongly constrain the difference in masses
between sparticles, but that the overall mass scale might only be
known to a precision of about 10\%. 
To determine the effect the sparticle spectrum
has on the SUSY comparison distribution, the masses of all of the sparticles
were simultaneously raised and lowered by $20~$GeV.
The change in the {\tt S5} comparison distribution, indicted by the
shaded band in \figref{fig:s5comparison} was small compared to the
statistical uncertainty, so we can conclude that, 
for this point, lack of knowledge of the SUSY mass scale will not
be a significant source of systematic uncertainty.
A similar test was not performed for the other points, but
it can be observed (\figref{fig:many-plots}) 
that the shape of the \costhetall\ distributions is rather universal, 
despite the differences in the spectra.

\subsection{Migrations at cut boundaries}
\label{sec:sys:cuts}
\EPSFIGURE{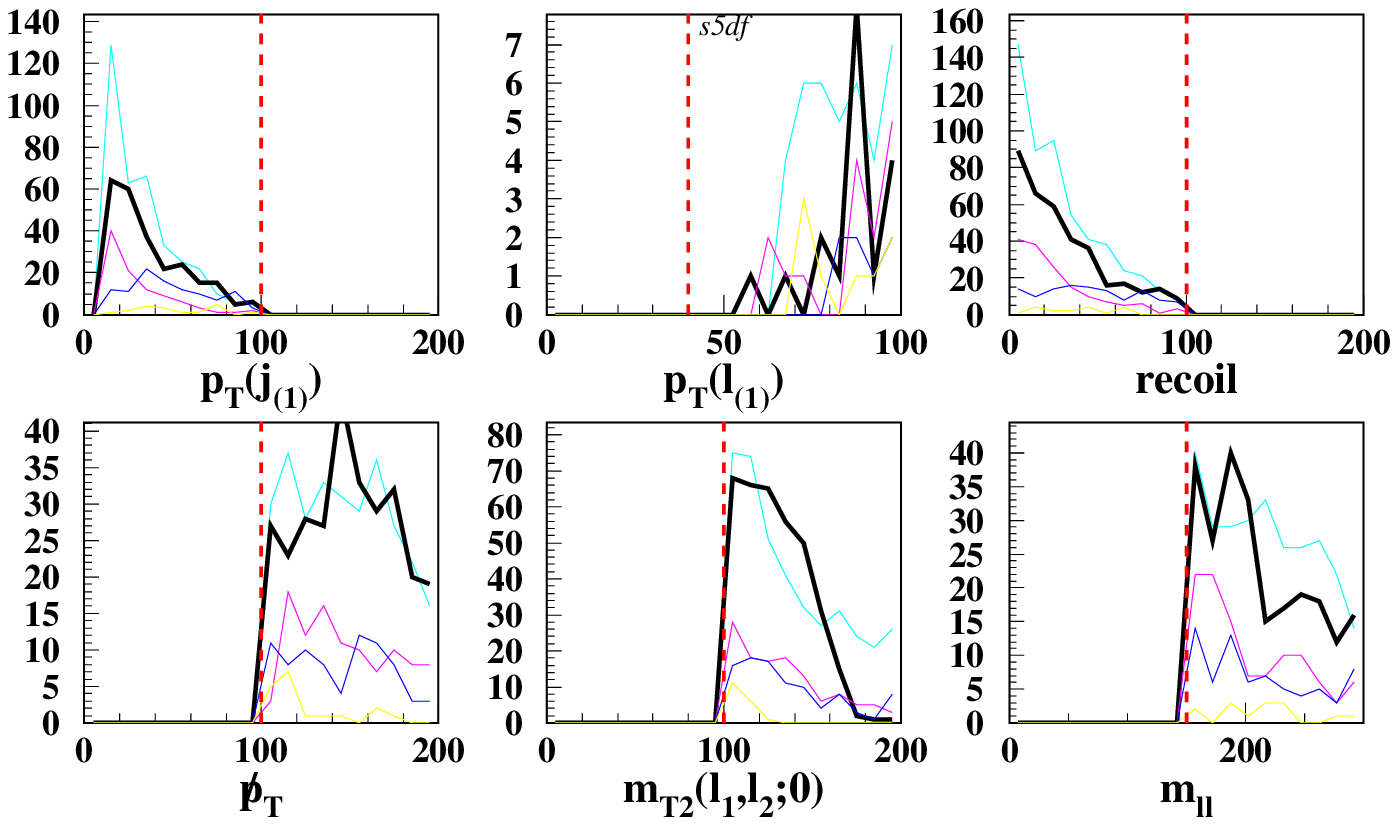, width=13cm}{
Distributions showing the number of events close to each of the various cuts,
for signal (thick line) and backgrounds (thin lines, not to scale).
The variables are the same selection
as those described in \secref{sec:cuts}.
Note that the distributions have been made {\em after} the cuts have been applied. 
The position of the cut is indicated by the vertical dashed line.
\label{fig:cuts}
}

If the distribution of interest has a strong dependance on a poorly-known
selection variable, this can generate a systematic uncertainty.
Migrations might occur, for example, from 
uncertainty in the hadronic energy scale, or from larger jet- or missing-energy 
resolution due to pile-up (minimum bias) events.
In \figref{fig:cuts} the number of signal and SM background events 
(not to scale) are plotted as a funtion of the selection variables.
The leading jet \pt, leading lepton \pt, and recoil have very few events near
their corresponding cut, so migrations due to uncertainties in these
variables should not significantly affect the final distributions.

The leading jet \pt\ and recoil cuts are 
about a factor of two larger than the value ($\approx 40$~GeV) at which 
pile-up events are expected to cause fake vetos at the percent 
level\cite{phystdr}.
The number of signal events lost by jets coming from 
pile-up events should therefore be negligible in this analysis.

The three variables on the bottom line of \figref{fig:cuts}, 
\ptmiss, \mttwo, and $m_{\ell\ell}$, {\em do} have 
a significant fraction of events close to their respective cuts, 
so a more careful study of the resolution in these variables would be appropriate.
For this analysis one is particularly interested in
finding out to what extent migrations might
change the {\em shape} of the \costhetall\ distribution.

While a detailed determination of the systematic uncertainties
requires a more complete detector simulation, and is 
beyond the scope of this paper,
we suggest that we might already have some optimism about the results.
A large proportion of the visible transverse energy in the signal 
events is carried by the final state leptons,
which should be well-measured in the electromagnetic calorimeter
(electrons) or in the tracking system (muons). This means that
the missing energy should also be relatively well measured,
and since all three of the variables $m_{ll}$, \ptmiss, and \mttwo,
depend only on $\Pt_\ell$, and \ptmiss, they 
ought to be well under control.

\subsection{Migrations between bins}
\label{sec:sys:bins}


Our bin size of $\Delta\costhetall=0.125$ 
corresponds to pseudorapidity differences of $\deltaetalep \geq 0.25$.
By contrast the $\eta$ segmentation in the first sampling of the 
ATLAS liquid argon calorimeter in the 
$|\eta|<2.5$ region is in the range 
0.003 to 0.006, depending on rapidity.
This fine $\eta$ segmentation will provide
a much better resolution than is required for our spin measurement. 
Similarly, for the muon reconstruction, the rapidity measurement
from the combined inner tracker and muon chambers will
be much more precise than is required here --
inter-bin migrations should not cause problems.

\subsection{Summary of systemic uncertainties}
\label{sys:concl}
In this section we have explored the various experimental 
systematic uncertainties which could potentially
reduce the power of the \costhetall\ distribution in making a spin measurement.
We have found that, while there are some areas -- such as
in SM background estimation from data, and in understanding the effect 
of the missing energy resolution -- which warrant further attention,
there is no reason to believe there are any serious problems - 
no `show-stoppers' which could invalidate the method.

\section{Conclusions}
\label{sec:conclusions}

We have described a new method for slepton spin determination at hadron colliders.
It is based on a measurement of the  di-slepton production anguar distribution, 
using a longitudinally boost-invariant variable we have called \costhetall.

The method is rather general, and is able to measure the spin 
for variety of supersymmetric model points.

We have examined LHC test point 5 ({\tt S5}), and seven of the Snowmass test 
points ({\tt SPS1a}$\to${\tt SPS6}).
Generally the results were very encouraging, though some
of the test points presented particular problems.
The method was unable to make a spin measurement at 
{\tt SPS2} (focus point region) because the production cross-section was 
much too small, nor at {\tt SPS6} (non-universal point) 
because the difference in mass between the sleptons and the lightest
neutralino is small. {\tt SPS4} is a difficult case, with relatively 
heavy sleptons, and a large SUSY backround.

Each of the other five points tested --
LHC point 5, {\tt SPS1a}, {\tt SPS1b}, {\tt SPS3}, and {\tt SPS 5} --
which include cosmological `bulk' points,
a `stau co-annihilation' point, and a point with a light \sstone\ 
allowed slepton spin determination
with integrated luminosity in the range 100 to 300~\invfb.

There do not appear to be any systematic uncertainties which 
might be seriously detrimental to the measurement, but 
we indicate that more work does need to be done in this area.

\section*{Acknowledgments}
Thank you to my colleagues in the UCL and Cambridge HEP groups, 
and to Shoji Asai, Giacomo Polesello, Peter Richardson and Dan Tovey
for useful comments and suggestions. 
I have made use of the physics analysis framework and tools which are
the result of ATLAS collaboration-wide efforts.
This work was supported by the Particle Physics and Astronomy Research Council.

\bibliography{thesis}

\end{document}

%% file: defs.tex
\newcommand{\newc}{\newcommand}
\newc{\definmath}[2] {\def#1{\ifmmode#2\else$#2$\fi}}

\definmath\gsim{\,\,\rlap{\raise 3pt\hbox{$>$}}{\lower 3pt\hbox{$\sim$}}\,\,}
\definmath\lsim{\,\,\rlap{\raise 3pt\hbox{$<$}}{\lower 3pt\hbox{$\sim$}}\,\,}

\definmath\amin{\mathrm{min}}
\definmath\amax{\mathrm{max}}

\def\compProg{\tt}
\def\herwig{{\compProg HERWIG}}
\def\herwigv#1{{\compProg HERWIG-#1}}

\def\isajetv#1{{\compProg ISAJET-#1}}

\def\atlfast{{\compProg ATLFAST}}
\def\atlfastv#1{{\compProg ATLFAST-#1}}

\newc{\barr}{\begin{eqnarray}}
\newc{\earr}{\end{eqnarray}}
\newc{\beq}{\begin{equation}}
\newc{\eeq}{\end{equation}}

\newc{\voidcol}{\phantom{m}\begin{rotate}{270}Not
reconstructed\end{rotate}\phantom{mn}}

\definmath\half{{\frac 1 2}}
\definmath\threehalfs{{\frac 3 2}}
\definmath\quarter{{\frac 1 4}}
\definmath\sixth{{\frac 1 6}}
\definmath\third{{\frac 1 3}}
\definmath\twothirds{{\frac 2 3}}
\definmath\fourthirds{{\frac 4 3}}
\definmath{\mPl}{M_\mathrm{Pl}}
\definmath{\invfb}{{\mathrm{fb}^{-1}}}
\definmath{\invab}{{\mathrm{ab}^{-1}}}
\definmath{\nsec}{\mathrm{ns}}

\definmath{\Omegadm}{\Omega_\mathrm{CDM}}
\definmath{\omegadm}{\omega_\mathrm{CDM}}

\definmath{\micron}{{\mu\mathrm{m}}}

\definmath{\silica}{{\mathrm{SiO_2}}}

\definmath{\mnought}{{m_0}}
\definmath{\mhalf}{{m_\frac{1}{2}}}
\definmath{\mthreehalfs}{{m_{3/2}}}
\definmath{\DeltaMChi}{{\Delta M_{\cht_1}}}
\definmath\mtx{{m_{TX}}}
\definmath\mttwo{{m_{T2}}}
\definmath\to{\rightarrow}
\newc{\mr}{\mathrm}

\def\slashchar#1{\setbox0=\hbox{$#1$}           
   \dimen0=\wd0                                 
   \setbox1=\hbox{/} \dimen1=\wd1               
   \ifdim\dimen0>\dimen1                        
      \rlap{\hbox to \dimen0{\hfil/\hfil}}#1 
   \else                                        
      \rlap{\hbox to \dimen1{\hfil$#1$\hfil}}/                                    \fi}

\definmath{\etmiss}{\slashchar{E}_T}
\definmath{\pmiss}{\slashchar{p}}
\definmath{\ptmiss}{\slashchar{p}_T}
\definmath{\Ptmiss}{\slashchar{{\bf p}}_T}
\definmath{\pt}{p_T}
\definmath{\Pt}{{{\bf p}_T}}
\definmath{\qth}{Q_\mr{thr}}

\newc{\appref}[1]{appendix~\ref{#1}}
\newc{\chref}[1]{chapter~\ref{#1}}
\newc{\Chref}[1]{Chapter~\ref{#1}}
\newc{\secref}[1]{section~\ref{#1}}
\newc{\secsref}[1]{sections~\ref{#1}}
\newc{\eqref}[1]{eq.~\ref{#1}}
\newc{\eqsref}[1]{eqs.~\ref{#1}}
\newc{\tabref}[1]{table~\ref{#1}}
\newc{\figref}[1]{fig.~\ref{#1}}
\newc{\figsref}[1]{figs.~\ref{#1}}
\newc{\partref}[1]{part~\ref{#1}}
\newc{\Secref}[1]{Section~\ref{#1}}
\newc{\Eqref}[1]{Eq.~\ref{#1}}
\newc{\Tabref}[1]{Table~\ref{#1}}
\newc{\Figref}[1]{Fig.~\ref{#1}}
\newc{\Partref}[1]{Part~\ref{#1}}

\definmath{\z}  {\mathrm{Z}^{0}}
\definmath\tbar{{\bar t}}
\definmath\ttbar{{t \tbar}}

\definmath{\cht}{\tilde{\chi}}
\definmath{\chgone}{{\cht^+_1}}
\definmath{\chgtwo}{{\cht^+_2}}
\definmath{\chgonem}{{\cht^-_1}}
\definmath{\chgonepm}{\cht^{\pm}_1}
\definmath{\chgall}{\cht^{\pm}_{1,2}}
\definmath{\ntlone}{{{\cht^0_1}}}
\definmath{\ntltwo}{{{\cht^0_2}}}
\definmath{\ntlthree}{\cht^0_3}
\definmath{\ntlfour}{\cht^0_4}
\definmath{\ntlall} {\tilde{\chi}_{1,2,3,4}^{0}}
\definmath{\gluino}{\tilde{g}}

\definmath{\ssul} {{\tilde{u}_{L}}}
\definmath{\ssdl} {{\tilde{d}_{L}}}
\definmath{\sscl} {\tilde{c}_{L}}
\definmath{\sssl} {\tilde{s}_{L}}
\definmath{\sstone} {\tilde{t}_{1}}
\definmath{\ssbone} {\tilde{b}_{1}}
\definmath{\ssur} {\tilde{u}_{R}}
\definmath{\ssdr} {\tilde{d}_{R}}
\definmath{\sscr} {\tilde{c}_{R}}
\definmath{\sssr} {\tilde{s}_{R}}
\definmath{\ssttwo} {\tilde{t}_{2}}
\definmath{\ssbtwo} {\tilde{b}_{2}}
\definmath{\squark} {{\tilde{q}}}
\definmath{\sqr} {\tilde{q}_{R}}
\definmath{\sql} {{\tilde{q}_{L}}}
\definmath{\squark} {{\tilde{q}}}
\definmath{\ssulbr} {\bar{\tilde{u}_{L}}}
\definmath{\ssdlbr} {\bar{\tilde{d}_{L}}}
\definmath{\ssclbr} {\bar{\tilde{c}_{L}}}
\definmath{\ssslbr} {\bar{\tilde{s}_{L}}}
\definmath{\sstonebr} {\bar{\tilde{t}_{1}}}
\definmath{\ssbonebr} {\bar{\tilde{b}_{1}}}

\definmath{\ssurbr} {\bar{\tilde{u}_{R}}}
\definmath{\ssdrbr} {\bar{\tilde{d}_{R}}}
\definmath{\sscrbr} {\bar{\tilde{c}_{R}}}
\definmath{\sssrbr} {\bar{\tilde{s}_{R}}}
\definmath{\ssttwobr} {\bar{\tilde{t}_{2}}}
\definmath{\ssbtwobr} {\bar{\tilde{b}_{2}}}

\definmath{\ssel} {\tilde{e}_{L}}
\definmath{\ssellp} {\tilde{e}_{L}^{+}}
\definmath{\ssellm} {\tilde{e}_{L}^{-}}
\definmath{\ssellpm} {\tilde{e}_{L}^{\pm}}

\definmath{\sser} {\tilde{e}_{R}}
\definmath{\sselrp} {\tilde{e}_{R}^{+}}
\definmath{\sselrm} {\tilde{e}_{R}^{-}}
\definmath{\sselrpm} {\tilde{e}_{R}^{\pm}}

\definmath{\ssmulp} {\tilde{\mu}_{L}^{+}}
\definmath{\ssmulm} {\tilde{\mu}_{L}^{-}}
\definmath{\ssmulpm} {\tilde{\mu}_{L}^{\pm}}

\definmath{\ssmurp} {\tilde{\mu}_{R}^{+}}
\definmath{\ssmurm} {\tilde{\mu}_{R}^{-}}
\definmath{\ssmurpm} {\tilde{\mu}_{R}^{\pm}}

\definmath{\sstauone} {{\tilde{\tau}_{1}}}
\definmath{\sstauonep} {\tilde{\tau}_{1}^{+}}
\definmath{\sstauonem} {\tilde{\tau}_{1}^{-}}
\definmath{\sstauonepm} {\tilde{\tau}_{1}^{\pm}}

\definmath{\sstautwop} {\tilde{\tau}_{2}^{+}}
\definmath{\sstautwom} {\tilde{\tau}_{2}^{-}}
\definmath{\sstautwopm} {\tilde{\tau}_{2}^{\pm}}

\definmath{\sstaupm} {\tilde{\tau}^{\pm}}
\definmath{\sstaump} {\tilde{\tau}^{\mp}}

\definmath{\sslrpm} {{\tilde{l}_{R}^{\pm}}}
\definmath{\sslr} {{\tilde{l}_{R}}}
\definmath{\ssll} {{\tilde{l}_{L}}}

\definmath{\ssnu} {\tilde{\nu}}
\definmath{\ssnuel} {\tilde{\nu}_{e}}
\definmath{\ssnumul} {\tilde{\nu}_{\mu}}
\definmath{\ssnutl} {\tilde{\nu}_{\tau}}

\definmath{\lqnear} {{l^\mathrm{near}q}}
\definmath{\lqfar} {{l^\mathrm{far}q}}
\definmath{\lqhigh} {{l^\mathrm{high}q}}
\definmath{\lqlow} {l{^\mathrm{low}q}}

\definmath{\lqbnear} {{l^\mathrm{near}\bar{q}}}
\definmath{\lqbfar} {{l^\mathrm{far}\bar{q}}}
\definmath{\lqbhigh} {{l^\mathrm{high}\bar{q}}}
\definmath{\lqblow} {{l^\mathrm{low}\bar{q}}}

\definmath{\mlqnear}{{m_{lq}^\mathrm{near}}}
\definmath{\mlqnearsq}{{\left(m_{lq}^\mathrm{near}\right)^2}}
\definmath{\mlqnearmax}{{\left(\mlqnear\right)_\mathrm{max}}}
\definmath{\mlqnearmaxsq}{{\left(\mlqnear\right)_\mathrm{max}^2}}

\definmath{\mlqbnear}{{m_{l\bar{q}}^\mathrm{near}}}
\definmath{\mlqfar}{{m_{lq}^\mathrm{far}}}
\definmath{\mlqbfar}{{m_{l\bar{q}}^\mathrm{far}}}

\definmath{\lqplus} {{l^+q}}
\definmath{\lqminus} {{l^-q}}

\definmath{\ww}{{W^+ W^-}}
\definmath{\zz}{{Z^0 Z^0}}
\definmath{\wz}{{W^\pm Z^0}}
\definmath{\ttbar}{{t \bar{t}}}
\definmath{\zgamma}{{Z^0 \gamma^*}}

\definmath{\lamp}{\lambda^\prime}
\definmath{\lampp}{\lambda^{\prime \prime}}
\definmath{\rparity}{{R_P}}

\newif\iftth
\iftth\else
\newc{\prepareAbbrev}[7]{\newcounter{#5}\newcommand{#1}{\mygloss{#2}{#6}{#4 #7}\ifnum\arabic{#5}=0 {#4 (#3)}\else#3\fi\addtocounter{#5}{1}}}
\fi